# A further step forward in measuring journals' scientific prestige: The SJR2 indicator

Vicente P. Guerrero-Bote[a] and Félix Moya-Anegón[b].

[a]University of Extremadura, Department of Information and Communication, Scimago Group, Spain.

[b]CSIC, CCHS, IPP, Scimago Group Spain.

**Abstract**

A new size-independent indicator of scientific journal prestige, the SJR2 indicator, is proposed. This indicator takes into account not only the prestige of the citing scientific journal but also its closeness to the cited journal using the cosine of the angle between the vectors of the two journals' cocitation profiles. To eliminate the size effect, the accumulated prestige is divided by the fraction of the journal's citable documents, thus eliminating the decreasing tendency of this type of indicator and giving meaning to the scores. Its method of computation is described, and the results of its implementation on the Scopus 2008 dataset is compared with those of an ad hoc Journal Impact Factor, JIF(3y), and SNIP, the comparison being made both overall and within specific scientific areas. All three, the SJR2 indicator, the SNIP indicator and the JIF distributions, were found to fit well to a logarithmic law. Although the three metrics were strongly correlated, there were major changes in rank. In addition, the SJR2 was distributed more equalized than the JIF by Subject Area and almost as equalized as the SNIP, and better than both at the lower level of Specific Subject Areas. The incorporation of the cosine increased the values of the flows of prestige between thematically close journals.

## 1. Introduction

It is accepted by the scientific community that neither all scientific documents nor all journals have the same value[1]. Instead of each researcher assigning a subjective value to each journal, there has always been strong interest in determining objective valuation procedures. In this regard, it is accepted by the scientific community that, in spite of different motivations (Brooks, 1985), citations constitute recognition of foregoing work (Moed, 2005).

One of the first generation of journal metrics based on citation counts developed to evaluate the impact of scholarly journals is the Impact Factor which has been extensively used for more than 40 years (Garfield, 2006). Nevertheless, different research fields have different yearly average citation rates (Lundberg, 2007), and this type of indicator is almost always lower in the areas of Engineering, Social Sciences, and Humanities (Guerrero et al., 2007; Lancho-Barrantes, Guerrero-Bote & Moya-Anegón, 2010a, 2010b).

Since neither all documents nor all journals have the same value, a second generation of indicators emerged with the idea of assigning them different weights. Rather than an index of popularity, the concept that it was intended to measure was prestige in the sense of Bonacich (1987) that the most prestigious journal will be the one that is most cited by journals also of high prestige. The first proposal in this sense in the field of Information Science was put forward by Pinski & Narin (1976), with a metric they called *"Journal Influence"*. With the arrival of the PageRank algorithm (Page et al., 1998) developed by the creators of Google, there have arisen other metrics such as the Invariant Method for the Measurement of Intellectual Influence (Palacios-Huerta & Volij, 2004), the Journal Status (Bollen, Rodríguez & van de Sompel, 2006), the Eigenfactor (Bergstrom, 2007), and the Scimago Journal Rank (González-Pereira, Guerrero-Bote & Moya-Anegón, 2010).

Despite the progress represented by this second generation of indicators, they have some features that make them ill-suited for journal metrics:

---

[1] Throughout this work, the term *"journal"* will be used indistinctly to refer to all the source publications in Scopus database for which the indices were calculated.

- The scores obtained by scientific journals typically represent their prestige, or their average prestige per document, but this score only makes sense in comparison with the scores of other journals.
- The scores are normalized by making them sum to a fixed quantity (usually, unity). The result is that as the number of journals increases the scores tend to decrease, which can lead to sets of indicators that all decrease with time. This characteristic complicates the study of the temporal evolution of scientific journals.
- Different scientific areas have different citation habits, and these are not taken into account in these indices, so that neither are the values obtained in different areas comparable (Lancho-Barrantes, Guerrero-Bote & Moya-Anegón, 2010b). Added to this is that there is no consensus on the classification of scientific journals into different areas (Janssens et al., 2009).

In the sciences, it has always been accepted that peer review in a field should be by experts in that same field (Kostoff, 1997). In this same sense, it seems logical to give more weight to citations from journals of the same or similar fields, since, although all researchers may use some given scientific study, they do not all have the same capacity to evaluate it. Even the weighting itself may not be comparable between different fields. Given this context, in a process of continuing improvement to find journal metrics that are more precise and more useful, the SJR2 indicator was designed to weight the citations according to the prestige of the citing journal, also taking into account the thematic closeness of the citing and the cited journals. The procedure does not depend on any arbitrary classification of scientific journals, but uses an objective informetric method based on cocitation. It also avoids the dependency on the size of the set of journals, and endows the score with a meaning that other indicators of prestige do not have.

In the following sections, we shall describe the methodological aspects of the development of the SJR2 indicator, and the results obtained with its implementation on Elsevier's Scopus database, for which the data were obtained from the Scimago

Journal & Country Rank website, an open access scientometric directory with almost 19,000 scientific journals and other types of publication (2009).

## 2. Data

We used Scopus as the data source for the development of the SJR2 indicator because it best represents the overall structure of world science at a global scale. Scopus is the world's largest scientific database if one considers the period 2000-2011. It covers most of the journals included in the Thomson Reuters Web of Science (WoS) and more (Moya-Anegón et al., 2007; Leydesdorff, Moya-Anegón & Guerrero-Bote, 2010). Also, despite its only relatively recent launch in 2004, there are already various studies of its structure and coverage in the literature (LaGuardia, 2005; Bar-Ilan, 2008; Jacso 2009). Our choice of database reflects our consideration of four criteria that are of great importance in the computation of any bibliometric indicator. These are:

- Journal coverage.
- Relationship between primary (citable items) and total output per journal of the database.
- Assignment criteria for types of documents.
- Accuracy of the linkage between references and source records.

Only documents published in 2008 included in the Scopus raw data copy exported on May 2011 were used for the main part of the study (in number, 1,999,777). All their references to documents present in the database in previous years were retrieved (in number, 26,036,560).

Documents are classified by area and category. There are 295 Specific Subject Areas grouped into 26 Subject Areas. In addition, there is the General Subject Area containing multidisciplinary journals, such as Nature or Science. The Subject Areas are grouped into four categories on the Scopus *"Basic Search"* page (see the Scopus website, www.scopus.com, visited on 20 October 2011).

The four Scopus categories are:

- Life Sciences (3950 titles): Agricultural & Biological Sciences; Biochemistry, Genetics & Molecular Biology; Immunology & Microbiology; Neuroscience, Pharmacology, Toxicology & Pharmaceutics.
- Physical Sciences (6350 titles): Chemical Engineering; Chemistry; Computer Science; Earth & Planetary Science; Energy; Engineering; Environmental Science; Materials Science; Mathematics; Physics & Astronomy.
- Social Sciences (5900 titles):Arts & Humanities; Business, Management & Accounting; Decision Sciences; Economics, Econometrics and Finance; Psychology; Social Sciences.
- Health Sciences (6200 titles, including 100% coverage of Medline titles): Medicine; Nursing; Veterinary; Dentistry; Health Professions.

## 3. Method

The SJR2 indicator, as also the SJR indicator (González-Pereira, Guerrero-Bote & Moya-Anegón, 2010), is computed over a journal citation network in which the nodes represent the active source journals, and the directed links between the nodes, the citation relationships among those journals. The main differences with respect to SJR are:

- The SJR2 prestige of the citing journal is distributed among the cited journals proportionally both to the citations from the former to the latter (in the three-year citation window) and to the cosine (of the angle) between the cocitation profiles of the two journals. With the addition of the cosine here, the intention is that the transfer should be greater the closer the two journals are thematically.
- The transfer of prestige to another journal or to itself is limited to a maximum of 50% of the prestige of the journal source, and a maximum of 10% per citation. This avoids problems similar to link farms with journals with either very few recent references or too specialized.

- The SJR2 prestige of the dangling nodes is distributed among all the journals proportionally to what they receive from the citing journals, which seems more logical than proportionally to the number of citable documents.
- The Prestige SJR2 (PSJR2) is normalized to the proportion of citable documents (articles, reviews, short surveys and conference papers in the three-year window), instead of to the total number of citable documents. With this, one obtain values that do not tend to decrease as new journals are incorporated and that are endowed with meaning.
- Short surveys have been included among the citable documents due to the non-negligible citation received by them[2].

The SJR2 indicator, as also the SJR, is computed in two phases: the computation of the Prestige SJR2 (PSJR2), a size-dependent measure that reflects the journals' overall prestige; and the normalization of this measure to give a size-independent metric, the SJR2 indicator, which can be used to compare journals.

**Phase 1**

First, each journal is assigned the same initial prestige value 1/N, where N is the number of journals in the database. Then the iterative procedure begins. Each iteration modifies the prestige values for each journal in accordance with three criteria: (1) a minimum prestige value from simply being included in the database; (2) a journal prestige given by the number of documents included in the database; and (3) a citation prestige given by the number, *"importance"*, and *"closeness"* of the citations received from other journals. The formula used for this calculation is the following:

$$PSJR2_i = \overbrace{\frac{(1-d-e)}{N}}^{1} + \overbrace{e \cdot \frac{Art_i}{\sum_{j=1}^{N} Art_j}}^{2} + \overbrace{\frac{d}{PSJR2D} \cdot \left[ \sum_{j=1}^{N} Coef_{ji} \cdot PSJR2_j \right]}^{3}$$

---

[2] The types of documents with a significant presence (> 1%) in Scopus in the citation window from 2005 to 2007 are: Article (64%) with 1.94 citations per document in 2008, Conference Papers (17%) with 0.49 c/d, Reviews (9%) with 2.47 c/d, Notes (2.45%) with 0.18 c/d, Editorial Material (2.29%) with 0.31 c/d, Letter (2.28%) with 0.36 c/d and Short Surveys (1.67%) with 0.76 c/d.

**PSJR2$_i$** – Prestige Scimago Journal Rank 2 of the Journal *i*.

**C$_{ji}$** - References from journal *j* to journal *i*.

**d** – Constant: 0.9.

**e** – Constant: 0.0999.

**N** - Number of journals in the database.

**Art$_j$** - Number of citable primary documents (articles, reviews, short surveys and conference papers) of journal *j*.

**Cos$_{ji}$** – Cosine between cocitation profiles of journals *j* and *i* (without components *i, j*).

The coefficients:

$$Coef_{ji} = \frac{(Cos_{ji} \cdot C_{ji})}{\sum_{h=1}^{N}(Cos_{jh} \cdot C_{jh})}$$

are calculated before beginning the iterations, and are limited to a maximun of *0.5* or *0.1·C$_{ji}$*. Unlike the SJR, in these coefficients the cosine of the cocitation profiles of the journals is introduced.

The factor:

$$PSJR2D = \sum_{i=1}^{N}\sum_{j=1}^{N} \frac{(Cos_{ji} \cdot C_{ji})}{\sum_{h=1}^{N}(Cos_{jh} \cdot C_{jh})} \cdot PSJR2_j$$

is calculated at the start of each iteration, and is the total PSJR2 distributed in that iteration (thus, with the PSJR2 of the dangling nodes not being included in the sum). Being the divisor, it provides the distribution of the PSJR2 of the dangling nodes, making the PSJR2 received by each journal grow proportionally until they all sum to unity, which without this factor would not be the case because of those dangling nodes. There was a similar correction factor, CF, in the SJR whose main purpose was to eliminate the difference between the active references used in the numerator of the coefficients and the total references used in the denominator, and which did not distribute the PSJR of dangling nodes.

The formula for the cosine of the cocitation profiles is:

$$Cos_{ij} = \frac{\sum_{h=1, h \neq i, h \neq j}^{N} Cocit_{ih} \cdot Cocit_{hj}}{\sqrt{\sum_{h=1, h \neq i, h \neq j}^{N} (Cocit_{ih})^2} \cdot \sqrt{\sum_{h=1, h \neq i, h \neq j}^{N} (Cocit_{jh})^2}}$$

**$Cocit_{ji}$** – Cocitation of journals *j* and *i*.

In which we do not include the cocitations between the two journals as these translate into differences since the self-cocitations of a journal are usually far more frequent than with other journals. For the calculation of the cocitation, only citations made in the year in question to the three-year window are used.

The scientific community accepts that the cocitations of documents (Marshakova, 1973; Small, 1973), authors (White, McCain, 1998), journals (McCain, 1991), and Subject Areas (Moya-Anegón et al., 2004) are indicators of the relationships among them. Thus, the cocitation between a pair of journals will indicate the relationship between them as a result of their having been used as sources in the same documents. But instead of using only the cocitation, the resolution is finer or more granular if one uses the cosine between the cocitation profiles. I.e., one not so much measures the direct relationship between two journals as the set of journals to which each is related in the sense that similar cocitation profiles will indicate a thematic relationship. We believe that it stands to reason that citations to scientific journals of related disciplines should have greater weight because of their greater capacity to evaluate a study, than citations to journals of very different disciplines. And it is then to be expected that this should have a normalizing effect on the various Subject Areas.

**Phase 2**

The *"Prestige SJR2"* (PSJR2) calculated in Phase 1 is a size-dependent metric that reflects the prestige of whole journals. It is not suitable for journal-to-journal comparisons since larger journals will tend to have greater prestige values. These values have the property of always summing to unity, so that they reflect the ratio of prestige that each scientific journal has accumulated. But, one needs to define a measure that is suitable for use in evaluation processes. To that end, the prestige gained by each journal, PSJR2, is divided by the ratio of citable documents that each journal has relative to the total, i.e.,

$$SJR2_i = \frac{PSJR2_i}{\left(Art_i \bigg/ \sum_{j=1}^{N} Art_j\right)} = \frac{PSJR2_i}{Art_i} \cdot \sum_{j=1}^{N} Art_j$$

The ratios of citable documents also have the characteristic of summing to unity. Hence this procedure compares the 'portion of the pie' of prestige that a journal achieves with the portion of citable documents that it includes. A value of unity means that the prestige per document is the mean. A value of 0.8 is interpreted as 20% less prestige having been achieved than the mean, and a value of 1.3 corresponds to 30% more prestige than the mean. Logically, an SJR2 value of 20 means that the prestige is 20 times greater than the mean.

Mathematically, it is easy to deduce that the mean of the SJR2 values for a year calculated by weighting by the number of citable documents will always be unity. In the SJR, since the divisor is just the number of articles of the journal, the scores decreased over time as a result of distributing a given measure of prestige among a growing number of journals. This was the contrary of the case with the JIF which grew as a result of the incorporation of ever more citations when further journals were incorporated.

Scopus distributes both the SJR and the SNIP (Source Impact Normalized per Paper) indicators. SNIP:

> "It measures a journal's contextual citation impact, taking into account characteristics of its properly defined subject field, especially the frequency at which authors cite other papers in their reference lists, the rapidity of maturing of citation impact, and the extent to which a database used for the assessment covers the field's literature" (Moed, 2010).

There is great variation from some subject fields to others in the database citation potential (number of references per document to the database and in the time period considered). To a large extent, this is the cause of the variation in citation impact from one subject field to another. One therefore normalizes the aforementioned citation impact, dividing it by the relative database citation potential (relative DCP) in the journal's subfield (the quotient between the DCP in the journal's subfield and the DCP of the database's median journal). Furthermore, to be classification free, the subject field used for each journal is the set of documents that cite its papers.

The SNIP indicator will also be used as a comparison point of the subject field normalization.

We have also constructed an ad hoc JIF(3y) with a 3-year citation window for comparison, so that any differences observed between the indicator values would be a consequence of the computation method and not of the time frame, citation window, etc.

Table 1 presents the main methodological differences with other indicators – the SNIP (Moed, 2010) and the JIF – and with other second generation prestige indicators – the Influence Weight (Pinski & Narin, 1976), Article Influence (Bergstrom, 2007), and the SJR itself (González-Pereira, Guerrero-Bote & Moya-Anegón, 2010).

Table 1. Methodological differences between the SJR2 indicator, SJR indicator, Article Influence, Influence Weight, SNIP and Impact Factor.

|  | SJR2 | SJR | Article Influence | Influence Weight | SNIP | Impact Factor |
|---|---|---|---|---|---|---|
| General differences | | | | | | |
| Source database | Scopus | Scopus | Web of Science | N.A. | Scopus | Web of Science |
| Citation time frame | 3 years | 3 years | 5 years | N.A. | 3 years | 2 years |
| Journal self-citation | Limited | Limited | Excluded | Included | Included | Included |
| Citation value | Weighted | Weighted | Weighted | Weighted | Unweighted | Unweighted |
| Size normalization | Citable document rate | Citable documents | Citable documents | Documents | Citable documents | Citable documents |
| Specific Influence Measures differences | | | | | | |
| Connection normalization | Normalized by the cosine weighted sum of active references in the citing journal | Normalized by the total number of references in the citing journal | Normalized by the number of active references in the citing journal | Normalized by the number of active references in the citing journal | N.A. | N.A. |
| Closeness weight | Cosine of cocitation profiles | N.A. | N.A. | N.A. | N.A. | N.A. |

## 4. Statistical Characterization

As in González-Pereira, Guerrero-Bote & Moya-Anegón (2010), in this section we shall present a statistical characterization of the SJR2 indicator in order to contrast its capacity to depict what could be termed *"average prestige"* with journals' citedness per document and the SNIP indicator. The study was performed for the year 2008 since its data can be considered stable. The data were downloaded from the Scimago Journal and Country Rank database (http://www.scimagojr.com) on 20 October 2011. It needs to be noted that while, due to the periodic SJR updates which include retrospective data, the data of the present study may not coincide exactly with those given on the portal, they will basically be the same.

Figure 1 shows a superposition of the overall SJR2, JIF(3y), and SNIP indicator values vs rank distributions. In order for them to be comparable, the values of the three indicators are normalized by dividing them by the corresponding maximum value. They all have a behaviour close to a logarithmic law which would be represented on this semi-log plot by a descending, although steeper, straight line. Contrary to the case with the SJR[3], SJR2 is now the indicator which has the most gradual fall, less steep even than the SNIP, with the JIF(3y) showing the sharpest decline. This indicates that the prestige is less concentrated than the Citation, i.e., that there are fewer *"prestigious"* journals than highly cited ones. The three metrics are strongly correlated. Relative to SJR[4], the SJR2 index has higher correlations with JIF(3y) and SNIP. There are also strong correlations with SNIP which are comparable to those between SNIP and

---

[3] With this set of data, SJR has also a somewhat steeper fall-off. The logarithmic approximation of the curve is *y= -0.017ln(x) + 0.1535* (i.e., smaller slope and closer to the x-axis) and its $R^2 = 0.4345$.

[4] With this set of data, the overall correlations between the SJR and the SJR2 were 0.794 (Pearson) and 0.863 (Spearman), between the SJR and the JIF(3y) 0.816 (Pearson) and 0.930 (Spearman), and between the SJR and the SNIP 0.454 (Pearson) and 0.731 (Spearman).
With this set of data, the mean correlations for Subject Areas between the SJR and the SJR2 were 0.781 (Pearson) and 0.916 (Spearman), between the SJR and the JIF(3y) 0.821 (Pearson) and 0.943 (Spearman), and between the SJR and the SNIP 0.630 (Pearson) and 0.827 (Spearman).
With this set of data, the mean correlations for Specific Subject Areas between the SJR and the SJR2 were 0.795 (Pearson) and 0.910 (Spearman), between the SJR and the JIF(3y) 0.815 (Pearson) and 0.917 (Spearman), and between the SJR and the SNIP 0.656 (Pearson) and 0.810 (Spearman).

JIF(3y). Table 2 gives details of these statistics, both overall and by Subject Area and Specific Subject Area.

Table 2: Overall correlations of the SJR2, JIF(3y), and SNIP indicators, and mean correlations by Subject Area and Specific Subject Area.

|  | Global | SJR2/JIF(3y) | SJR2/SNIP | SNIP/JIF(3y) |  |  |
|---|---|---|---|---|---|---|
|  | Pearson | 0.882 | 0.775 | 0.771 |  |  |
|  | Spearman | 0.944 | 0.906 | 0.888 |  |  |
| **Subject Areas (27)** | SJR2/JIF(3y) | | SJR2/SNIP | | SNIP/JIF(3y) | |
|  | Average | SD | Average | SD | Average | SD |
| Pearson | 0.910 | 0.072 | 0.868 | 0.105 | 0.912 | 0.064 |
| Spearman | 0.944 | 0.039 | 0.910 | 0.052 | 0.924 | 0.026 |
| **Specific Subject Areas (295)** | SJR2/JIF(3y) | | SJR2/SNIP | | SNIP/JIF(3y) | |
|  | Average | SD | Average | SD | Average | SD |
| Pearson | 0.873 | 0.241 | 0.842 | 0.213 | 0.872 | 0.208 |
| Spearman | 0.917 | 0.179 | 0.882 | 0.144 | 0.906 | 0.132 |

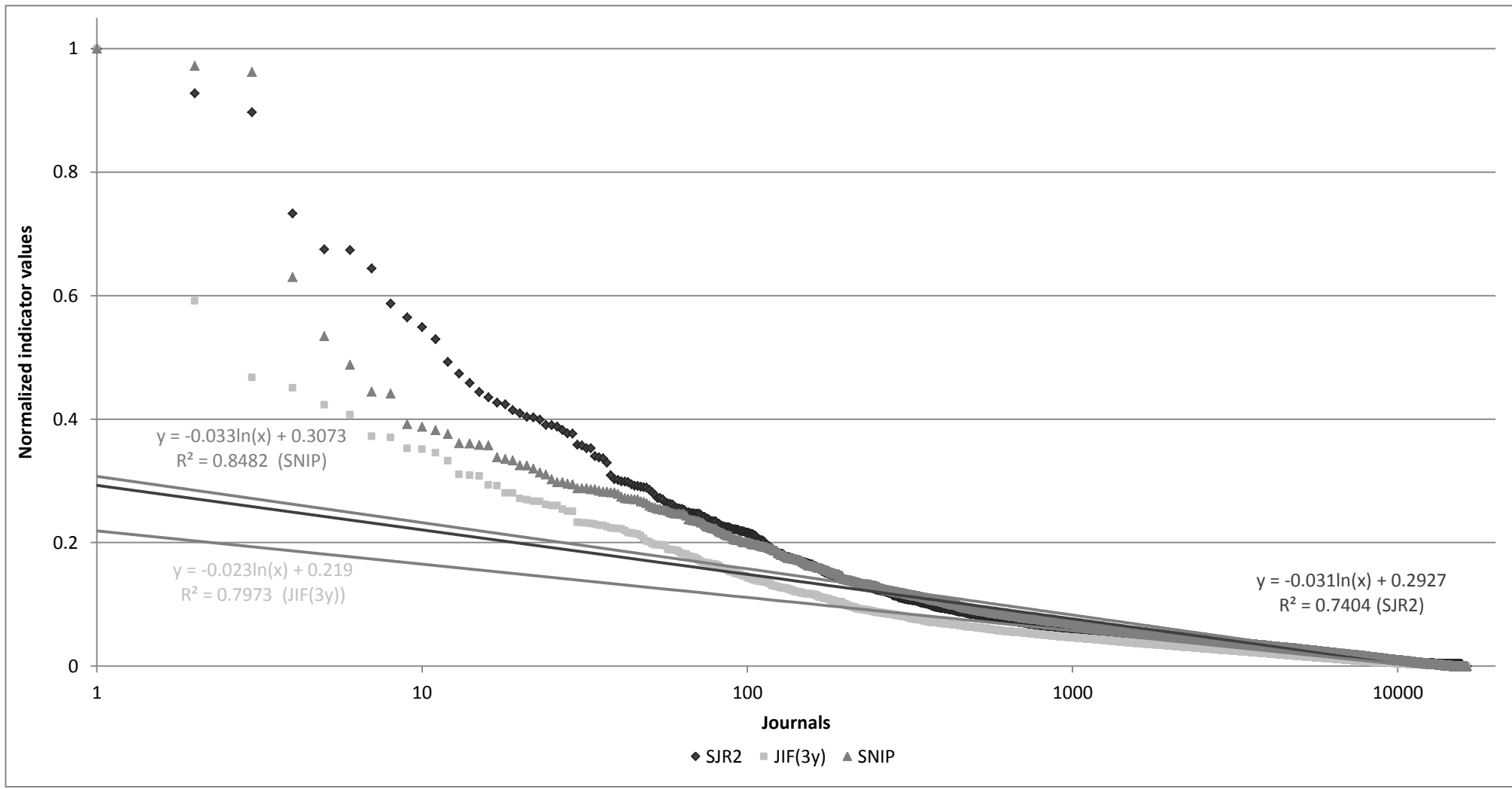

Figure 1: Superposition of the SJR2, SNIP, and JIF(3y) indicator values vs rank distributions (normalized by their respective maxima).

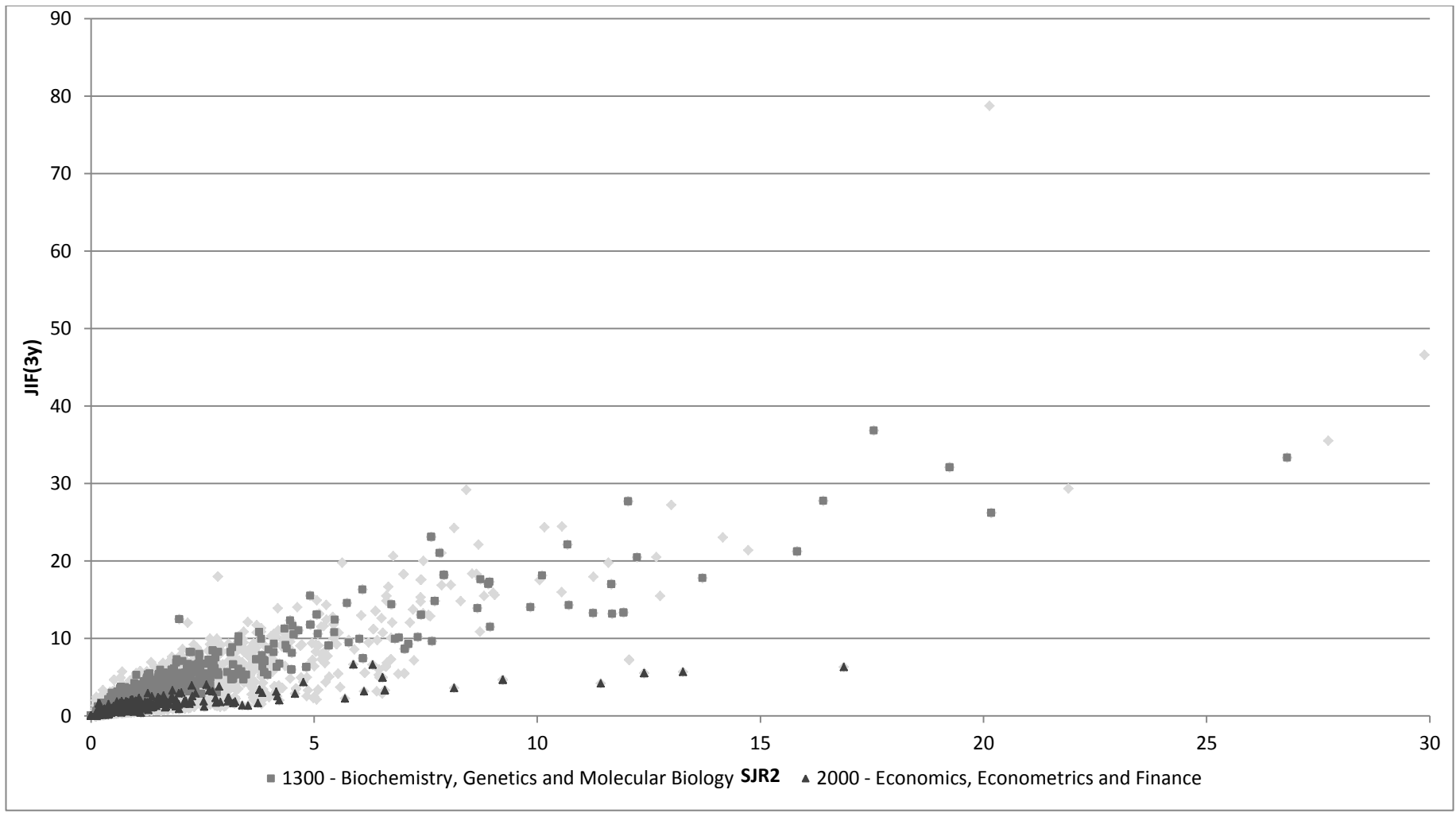

Figure 2: Scatterplot of JIF(3y) vs the SJR indicator. The Biochemistry, Genetics & Molecular Biology, and Economics, Econometrics & Finance Subject Areas are highlighted.

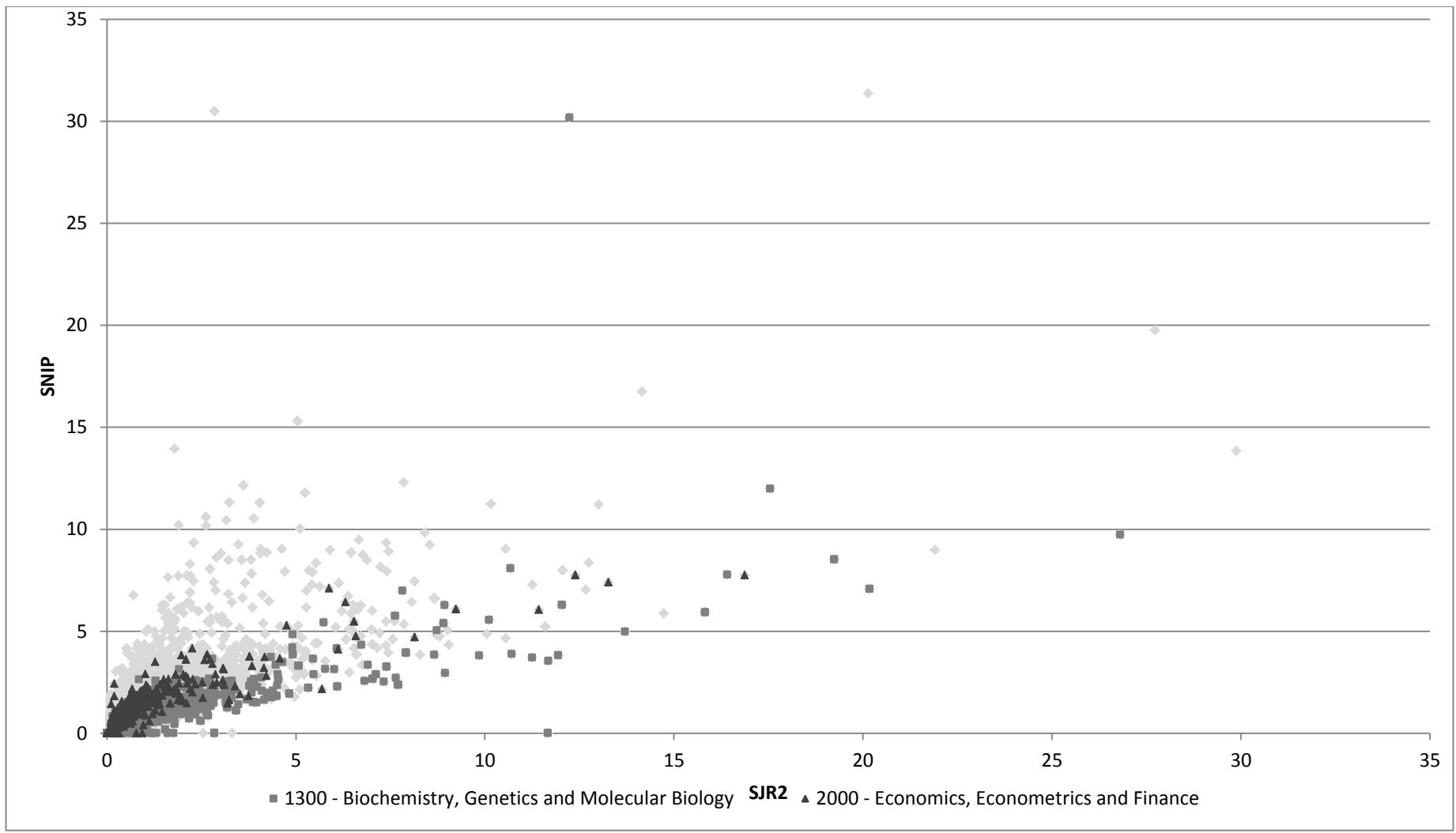

Figure 3: Scatterplot of SNIP vs the SJR2 indicator. The Biochemistry, Genetics & Molecular Biology, and Economics, Econometrics & Finance Subject Areas are highlighted.

Figures 2 and 3 are scatter-plots of the same distributions as shown in Figure 1. They show all the journals for which the SNIP and SJR indicators are currently estimated, but they also mark as highlighted two Subject Areas of very different behaviour in terms of the traffic of citations. In the first (Figure 2), which shows SJR2 vs JIF(3y), one observes the normalizing effect that SJR2 has on the different citation habits. The journals of the area *"1300 - Biochemistry, Genetics and Molecular Biology"* lie above those corresponding to *"2000 - Economics, Econometrics and Finance"* as a result of having higher JIF(3y) values. Indeed, one of the journals of the latter Subject Area with a modest impact of 6.29 obtains an outstanding SJR2 of 16.87.

Figure 3 shows the case to be the inverse with the SNIP, with the journal of *"2000 - Economics, Econometrics and Finance"* having SNIP values greater than those of *"1300 - Biochemistry, Genetics and Molecular Biology"*. This is perhaps because of an over-normalization of this indicator as a result of the computation being carried out solely by numerical comparison with citing journals.

This is seen numerically in Table 3 which lists the calculated citation rates in the different Subject Areas with respect to the cumulative total for each indicator, weighted by the number of citable documents of each journal. In the case of SJR2, this is the Prestige SJR2 (PSJR2). These values are divided by the ratio of citable documents of each Subject Area. Thus a situation of complete equalization should yield unity for each Subject Area.

Table 3: Subject Area distribution of the citation rates of the SJR2, JIF(3y), and SNIP indicators.

| Area | SJR2 | JIF(3y) | SNIP |
|---|---|---|---|
| General | 4.133 | 4.367 | 2.978 |
| Agricultural and Biological Sciences | 0.897 | 0.940 | 0.981 |
| Arts and Humanities | 0.230 | 0.130 | 0.344 |
| Biochemistry, Genetics and Molecular Biology | 1.683 | 1.855 | 1.184 |
| Business, Management and Accounting | 0.740 | 0.491 | 0.923 |
| Chemical Engineering | 0.712 | 0.802 | 0.875 |
| Chemistry | 1.195 | 1.369 | 1.116 |
| Computer Science | 0.805 | 0.606 | 1.446 |
| Decision Sciences | 1.139 | 0.698 | 1.690 |
| Earth and Planetary Sciences | 1.166 | 0.976 | 1.192 |
| Economics, Econometrics and Finance | 1.220 | 0.573 | 1.283 |
| Energy | 0.588 | 0.554 | 0.878 |
| Engineering | 0.641 | 0.516 | 1.067 |
| Environmental Science | 0.986 | 1.029 | 1.112 |
| Immunology and Microbiology | 1.561 | 1.810 | 1.241 |
| Materials Science | 0.817 | 0.788 | 0.917 |
| Mathematics | 0.837 | 0.494 | 1.003 |
| Medicine | 0.875 | 1.126 | 0.844 |
| Neuroscience | 1.955 | 2.106 | 1.357 |
| Nursing | 0.637 | 0.761 | 0.674 |
| Pharmacology, Toxicology and Pharmaceutics | 0.792 | 1.178 | 0.774 |
| Physics and Astronomy | 1.146 | 0.942 | 1.151 |
| Psychology | 0.928 | 0.933 | 1.094 |
| Social Sciences | 0.519 | 0.389 | 0.711 |
| Veterinary | 0.479 | 0.488 | 0.639 |
| Dentistry | 0.715 | 0.837 | 1.000 |
| Health Professions | 0.844 | 1.066 | 1.070 |

As expected, the values that most deviate from unity are those of the *"General"* Subject Area. But it must be borne in mind that this is a special Subject Area which includes multidisciplinary journals that publish work from practically any discipline, and, as one observes, accumulate a Citation close to four times unity. One also observes that the journals of this Subject Area obtain a somewhat higher PSJR2, indicating that their citations come from prestigious journals. The SNIP indicator is the one that least deviates from unity in this Subject Area.

Here one observes that *"1300 - Biochemistry, Genetics and Molecular Biology"* accumulates a greater Citation (1.8) than Prestige SJR2 (1.68) or SNIP (1.18), while *"2000 - Economics, Econometrics and Finance"* presents the opposite behaviour.

Table 4 summarizes the average squared deviations from unity for Subject Areas and for Specific Subject Areas. In neither case was the Subject Area *"1000 - General"* taken into account because of its aforementioned special nature. One observes in this table that the greatest deviation from unity corresponds to JIF(3y). In the case of the Subject Areas, the most equalized result is obtained by SNIP followed closely by SJR2, while for the Specific Subject Areas although SNIP also has the most equalized normalization, SJR2 is still closer[5].

Table 4: Mean squared deviation from unity of the distribution of the rates of the SJR2, JIF(3y), and SNIP, by Subject Area and by Specific Subject Area.

| Average of squared differences to the unity | SJR2 | JIF(3y) | SNIP |
|---|---|---|---|
| Subject Areas (26) | 0.146 | 0.221 | 0.075 |
| Specific Subject Areas (294) | 0.278 | 0.344 | 0.262 |

Table 5: Particular case of different values of the indicators for two journals.

| Journal | | ACM Computing Surveys | Foundations and Trends in Communications and Information Theory |
|---|---|---|---|
| Sourcerecord Id | | 23038 | 4000151805 |
| SJR2 | | 2.84 | 12.06 |
| JIF(3y) | | 17.97 | 7.2 |
| SNIP | | 30.49 | 7.98 |
| SJR | | 0.2661 | 0.2452 |
| Citable papers | | 36 | 10 |
| Total Citation | | 647 | 72 |
| SJR2 Considered Citation | | 336 | 54 |
| Average of cosine of citations | | 0.13 | 0.67 |
| Average of SJR2 of citations | | 0.92 | 2.07 |
| PSJR2 from citations | | 2.264E-05 | 2.741E-05 |
| PSJR2 | | 2.347E-05 | 2.765E-05 |
| First Contributor | Citations Considered | 101 | 7 |
| | Cosine | 0.132 | 0.909 |
| | SJR2 | 0.316 | 6.882 |
| | PSJR2 | 0.0051 | 0.0020 |
| | Contribution | 5.904E-06 | 1.318E-05 |
| | Papers (2008) | 19161 | 471 |
| | References (2008) | 327712 | 11988 |

By way of a case study, Table 5 lists the data for two journals with different values. These are two journals of the Subject Area *"1500 - Computer Science"*. They both have a high JIF(3y) in this Subject Area, although the value for the first of them is more than twice that of the second. With the SNIP normalization, both obtain higher values (than JIF(3y)), but now there is an enhancement of the difference which now reaches a factor of more than three. However, the order is completely reversed with SJR2, the second now having a value four times that of the first. In the table, one can see what the reasons are for this change. Firstly, the Total Citation which is used to calculate JIF(3y) and SNIP for the first journal is almost twice that considered in SJR2.

---

[5] The average squared deviations from unity of the SJR for Subject Areas and for Specific Subject Areas were greater than those shown: 0.584 and 0.762, respectively.

This is because many of the citations obtained by the first journal come from journals which, although they are included in Scopus, for different reasons are not included in the calculation of SJR. Secondly, the average cosine of the cocitation profiles of the source journals of the citations received by the second journal is five times that of the first journal. And thirdly, the average SJR2 of the source journals of the citations received by the second journal is more than double the first. Altogether, this leads to the Prestige SJR2 received from citations and the total Prestige SJR2 being some 20% higher in the second journal, while the number of citable documents is almost four times greater in the first journal.

The same table presents the data for the greatest contributor for each of the two journals. For the first journal, 101 citations are from a journal that has a considerable PSJR2, but a very low cosine value, and many references among which its PSJR2 is distributed, being a journal with 19,161 documents in 2008. For the second journal, there are only 7 citations from a journal with less than half the value of PSJR2, but a cosine of 0.9 and far fewer references among which to distribute its PSJR2 since it published only 471 documents in 2008. This leads to the second journal's greatest contribution being more than twice that of the first journal.

As mentioned above, the effect desired with the cosine between cocitation profiles is to give greater weight to the prestige from thematically related journals. This means that greater value will be given to the Citation from the same Subject Area or Specific Subject Area. This can be seen in Table 6 which lists the citation flow percentages of Prestige SJR2 with and without the cosine effect. One observes in the table how the citation habits of different Subject Areas vary from 17% of the Citation coming from the same Specific Subject Area in *"Nursing"* to 63% in *"Dentistry"*. One also observes that the SJR2 (even without the cosine) increases the value of the flow percentage from the same Subject Area and Specific Subject Area (except in the area "Agricultural and Biological Sciences", due mainly to the large prestige per citation in the special subject area "General" and to the large ratio of citation in "Economics, Econometrics and Finance"). This increase is greater when the cosine is included.

The increase in the rate of Subject Areas such as Decision Sciences (Table 3) can also be explained as due to the almost doubling of the flow within that Subject Area or its Specific Subject Areas.

The averages of these data are presented in Table 7, which also gives the percentages of self-citation flows and the percentages of outgoing flows. One sees in the table that, despite limiting consideration to self-citations, SJR2 increases the weight of the flow to or from the same journal. The increases are greater when the cosine is included. This was to be expected, since the cosine of a self-cocitation vector is unity, the highest possible value. The same is the case with the flows from the same Subject Area or Specific Subject Area.

To provide a general overview, the flows of Prestige SJR2 between Subject Areas are listed in Table 8 and shown graphically in Figure 4. If one were to generate the corresponding figure for the Citation, as well as the changes in grey levels of the nodes because the accumulated prestige is different from the accumulated citation, one would see how the thickness of the loops would decrease, while that of the links between classes would increase. This is because, in addition to taking into account the prestige of the source journal, SJR2, through cocitation profiles, it takes into account the thematic proximity between citing and cited journal.

Two clusters with high traffic of prestige can be distinguished in Figure 4. One of Biomedicine (includes the general area of multidisciplinary journals) and another of Physics, Chemistry and Engineering

Table 6: Percentage flows of Citation and Prestige SJR2 (with and without cosine effect) received from the same Subject Area or Specific Subject Area.

| Area | Specific Subject Area | | | Subject Area | | |
|---|---|---|---|---|---|---|
| | % Citation | % SJR2 (Without Cosine) | % SJR2 | % Citation | % SJR2 (Without Cosine) | % SJR2 |
| General | 4.95 | 19.17 | 30.69 | 4.95 | 19.17 | 30.69 |
| Agricultural and Biological Sciences | 40.60 | 37.04 | 52.64 | 57.32 | 51.64 | 63.67 |
| Arts and Humanities | 41.75 | 46.87 | 61.46 | 48.98 | 55.05 | 65.57 |
| Biochemistry, Genetics and Molecular Biology | 27.20 | 30.05 | 38.51 | 51.98 | 54.78 | 61.54 |
| Business, Management and Accounting | 38.37 | 48.31 | 63.66 | 60.63 | 63.88 | 76.24 |
| Chemical Engineering | 28.65 | 29.27 | 45.05 | 39.28 | 38.34 | 52.23 |
| Chemistry | 39.77 | 38.45 | 53.04 | 67.33 | 63.48 | 76.13 |
| Computer Science | 31.22 | 37.99 | 53.71 | 56.61 | 64.85 | 77.92 |
| Decision Sciences | 30.60 | 41.43 | 61.17 | 33.11 | 43.52 | 62.85 |
| Earth and Planetary Sciences | 53.41 | 58.74 | 67.84 | 74.04 | 77.61 | 86.93 |
| Economics, Econometrics and Finance | 49.46 | 68.58 | 77.83 | 59.30 | 76.19 | 84.53 |
| Energy | 30.70 | 37.20 | 54.79 | 37.47 | 44.58 | 60.76 |
| Engineering | 39.09 | 44.41 | 59.33 | 53.15 | 58.04 | 69.51 |
| Environmental Science | 34.69 | 36.22 | 52.15 | 46.15 | 46.43 | 61.06 |
| Immunology and Microbiology | 33.88 | 37.37 | 50.18 | 43.94 | 46.07 | 58.69 |
| Materials Science | 32.73 | 32.23 | 45.90 | 53.02 | 50.29 | 62.41 |
| Mathematics | 39.15 | 46.29 | 61.24 | 52.89 | 64.28 | 73.92 |
| Medicine | 32.66 | 33.37 | 49.76 | 70.56 | 67.91 | 74.78 |
| Neuroscience | 25.68 | 30.92 | 41.72 | 39.57 | 42.95 | 54.58 |
| Nursing | 17.45 | 18.53 | 31.12 | 23.20 | 21.96 | 34.73 |
| Pharmacology, Toxicology and Pharmaceutics | 23.74 | 20.25 | 33.21 | 32.64 | 26.61 | 39.29 |
| Physics and Astronomy | 39.74 | 42.58 | 55.79 | 62.88 | 66.08 | 75.90 |
| Psychology | 27.82 | 31.36 | 44.06 | 42.42 | 45.66 | 57.75 |
| Social Sciences | 32.69 | 38.61 | 58.32 | 50.21 | 56.33 | 69.84 |
| Veterinary | 38.66 | 36.69 | 59.03 | 52.58 | 48.77 | 72.00 |
| Dentistry | 63.10 | 64.79 | 85.91 | 65.39 | 65.39 | 86.54 |
| Health Professions | 17.79 | 22.33 | 39.77 | 18.69 | 23.05 | 40.31 |

Table 7: Averages, weighted by the number of citable documents, of the percentage flows of Citation and Prestige SJR2 (with or without cosine effect) received from or sent to the same journal, Subject Area, or Specific Subject Area, as calculated by Subject Area and by Specific Subject Area.

| | | | Subject Areas (27) | Specific Subject Areas (295) |
|---|---|---|---|---|
| Sent | Self | Journal Selfreferencing | 10.90 | 11.05 |
| | | Journal Self PSJR2 (wc) sent | 13.27 | 13.25 |
| | | Journal Self PSJR2 sent | 23.51 | 23.97 |
| | Specific Subject Area | Referencing inside Specific Subject Area | 32.63 | 29.10 |
| | | PSJR2 (wc) sent inside Specific Subject Area | 35.88 | 31.40 |
| | | PSJR2 sent inside Specific Subject Area | 49.63 | 44.83 |
| | Subject Area | Referencing in Subject Area | 53.45 | 55.56 |
| | | PSJR2 (wc) sent inside Subject Area | 55.99 | 57.13 |
| | | PSJR2 sent inside Subject Area | 65.99 | 66.99 |
| Received | Self | Journal Selfcitation | 11.65 | 12.08 |
| | | Journal Self PSJR2 (wc) received | 13.87 | 14.13 |
| | | Journal Self PSJR2 received | 24.64 | 25.77 |
| | Specific Subject Area | Citation from the same Specific Subject Area | 34.72 | 31.00 |
| | | PSJR2 (wc) received from the same Specific Subject Area | 37.29 | 33.12 |
| | | PSJR2 received from the same Specific Subject Area | 51.69 | 47.44 |
| | Subject Area | Citation from the same Subject Area | 56.66 | 58.56 |
| | | PSJR2 (wc) received from the same Subject Area | 57.98 | 59.69 |
| | | PSJR2 received from the same Subject Area | 68.50 | 70.18 |

**Table 8: Flow of Prestige SJR2 between the different Subject Areas.**

| | General | Agricultural and Biological Sciences | Arts and Humanities | Biochemistry, Genetics and Molecular Biology | Business, Management and Accounting | Chemical Engineering | Chemistry | Computer Science | Decision Sciences | Earth and Planetary Sciences | Economics, Econometrics and Finance | Energy | Engineering | Environmental Science | Immunology and Microbiology | Materials Science | Mathematics | Medicine | Neuroscience | Nursing | Pharmacology, Toxicology and Pharmaceutics | Physics and Astronomy | Psychology | Social Sciences | Veterinary | Dentistry | Health Professions |
|---|---|---|---|---|---|---|---|---|---|---|---|---|---|---|---|---|---|---|---|---|---|---|---|---|---|---|---|
| General | 0.0119 | 0.0029 | 1E-05 | 0.0142 | 1E-05 | 0.0001 | 0.0007 | 0.0002 | 6E-06 | 0.0008 | 2E-05 | 2E-05 | 0.0004 | 0.0006 | 0.0025 | 0.0003 | 0.0002 | 0.004 | 0.0017 | 1E-05 | 0.0004 | 0.0014 | 8E-05 | 5E-05 | 5E-06 | 1E-06 | 8E-06 |
| Agricultural and Biological Sciences | 0.0029 | 0.0433 | 5E-05 | 0.0063 | 3E-05 | 0.0002 | 0.0006 | 0.0002 | 1E-05 | 0.001 | 6E-05 | 7E-05 | 0.0002 | 0.0032 | 0.0012 | 0.0001 | 6E-05 | 0.0023 | 0.0007 | 0.0002 | 0.0006 | 8E-05 | 0.0001 | 0.0001 | 8E-05 | 5E-06 | 2E-05 |
| Arts and Humanities | 3E-05 | 7E-05 | 0.0018 | 4E-05 | 1E-05 | 1E-06 | 1E-05 | 5E-05 | 1E-06 | 3E-05 | 3E-05 | 6E-07 | 3E-05 | 2E-05 | 3E-06 | 3E-06 | 2E-05 | 0.0001 | 8E-05 | 3E-06 | 5E-07 | 2E-05 | 0.0002 | 0.0002 | 8E-08 | 4E-08 | 2E-05 |
| Biochemistry, Genetics and Molecular Biology | 0.0153 | 0.0054 | 2E-05 | 0.1286 | 1E-05 | 0.0004 | 0.0024 | 0.0001 | 2E-05 | 6E-05 | 5E-06 | 3E-05 | 0.0004 | 0.0007 | 0.0064 | 0.0004 | 0.0001 | 0.0108 | 0.0043 | 0.0006 | 0.0028 | 0.0005 | 0.0003 | 8E-05 | 0.0001 | 6E-05 | 0.0002 |
| Business, Management and Accounting | 2E-05 | 4E-05 | 1E-05 | 3E-05 | 0.0096 | 2E-05 | 6E-06 | 0.0003 | 0.0001 | 2E-05 | 0.0004 | 2E-05 | 0.0002 | 5E-05 | 2E-05 | 1E-05 | 0.0001 | 8E-05 | 7E-06 | 7E-06 | 7E-06 | 6E-06 | 0.0003 | 0.0004 | 2E-05 | 7E-08 | 2E-06 |
| Chemical Engineering | 0.0002 | 0.0003 | 1E-06 | 0.0005 | 7E-06 | 0.0125 | 0.0016 | 8E-05 | 6E-06 | 0.0002 | 5E-06 | 0.0002 | 0.0007 | 0.0004 | 7E-05 | 0.0011 | 0.0001 | 0.0001 | 2E-05 | 3E-05 | 0.0002 | 0.0011 | 8E-06 | 2E-05 | 2E-06 | 5E-06 | 4E-06 |
| Chemistry | 0.001 | 0.0005 | 1E-05 | 0.0032 | 1E-06 | 0.0013 | 0.0702 | 9E-05 | 5E-07 | 0.0001 | 7E-08 | 0.0002 | 0.0008 | 0.0005 | 0.0003 | 0.0044 | 0.0001 | 0.0006 | 8E-05 | 1E-05 | 0.0007 | 0.004 | 2E-06 | 3E-05 | 4E-06 | 3E-06 | 2E-05 |
| Computer Science | 0.0003 | 0.0002 | 6E-05 | 0.0003 | 0.0004 | 1E-04 | 0.0001 | 0.0347 | 0.0002 | 0.0001 | 7E-05 | 6E-05 | 0.0017 | 6E-05 | 6E-05 | 0.0002 | 0.0006 | 0.0003 | 0.0001 | 7E-06 | 4E-05 | 0.0008 | 4E-05 | 0.0002 | 6E-07 | 4E-07 | 2E-05 |
| Decision Sciences | 5E-06 | 1E-05 | 8E-07 | 2E-05 | 0.0002 | 3E-06 | 6E-07 | 0.0002 | 0.0037 | 2E-06 | 0.0002 | 4E-06 | 0.0001 | 7E-06 | 7E-07 | 2E-06 | 0.0002 | 2E-05 | 3E-07 | 2E-07 | 2E-07 | 2E-05 | 7E-06 | 2E-05 | 9E-08 | 1E-08 | 1E-06 |
| Earth and Planetary Sciences | 0.0008 | 0.0009 | 3E-05 | 0.0001 | 1E-05 | 0.0002 | 0.0002 | 8E-05 | 4E-06 | 0.0419 | 2E-05 | 8E-05 | 0.0004 | 0.0008 | 7E-05 | 0.0002 | 0.0001 | 8E-05 | 3E-06 | 1E-07 | 4E-05 | 0.0004 | 2E-06 | 6E-05 | 1E-06 | 8E-08 | 2E-07 |
| Economics, Econometrics and Finance | 1E-05 | 5E-05 | 3E-05 | 6E-06 | 0.0004 | 2E-06 | 2E-07 | 4E-05 | 0.0001 | 1E-05 | 0.012 | 8E-06 | 4E-05 | 7E-05 | 1E-06 | 8E-07 | 0.0002 | 9E-05 | 4E-06 | 7E-06 | 3E-06 | 4E-06 | 6E-05 | 0.0004 | 1E-07 | 2E-07 | 8E-06 |
| Energy | 2E-05 | 0.0001 | 7E-07 | 6E-05 | 2E-05 | 0.0003 | 0.0003 | 4E-05 | 3E-06 | 9E-05 | 1E-05 | 0.0042 | 0.0005 | 0.0002 | 2E-05 | 0.0003 | 1E-05 | 7E-05 | 2E-07 | 4E-08 | 2E-05 | 0.0004 | 2E-06 | 6E-05 | 7E-08 | 1E-08 | 2E-05 |
| Engineering | 0.0006 | 0.0002 | 3E-05 | 0.0006 | 0.0002 | 0.0008 | 0.0011 | 0.0017 | 0.0002 | 0.0005 | 7E-05 | 0.0006 | 0.0501 | 0.0002 | 9E-05 | 0.0023 | 0.0007 | 0.0005 | 0.0001 | 5E-05 | 0.0002 | 0.0053 | 5E-05 | 0.0002 | 2E-06 | 3E-05 | 5E-05 |
| Environmental Science | 0.0007 | 0.003 | 2E-05 | 0.0009 | 5E-05 | 0.0004 | 0.0006 | 5E-05 | 6E-06 | 0.0008 | 9E-05 | 0.0002 | 0.0001 | 0.0235 | 0.0003 | 1E-04 | 3E-05 | 0.0006 | 9E-05 | 1E-05 | 0.0003 | 0.0001 | 9E-06 | 0.0002 | 3E-05 | 2E-06 | 6E-06 |
| Immunology and Microbiology | 0.0026 | 0.001 | 5E-07 | 0.0067 | 9E-06 | 5E-05 | 0.0002 | 6E-05 | 2E-07 | 4E-05 | 1E-06 | 6E-06 | 7E-05 | 0.0003 | 0.031 | 2E-05 | 2E-05 | 0.0063 | 0.0004 | 5E-05 | 0.0004 | 7E-05 | 7E-06 | 6E-05 | 1E-04 | 3E-05 | 2E-05 |
| Materials Science | 0.0005 | 0.0001 | 3E-06 | 0.0005 | 3E-06 | 0.0012 | 0.0053 | 0.0002 | 2E-06 | 0.0001 | 1E-06 | 0.0003 | 0.0019 | 9E-05 | 4E-05 | 0.0319 | 0.0002 | 0.0001 | 9E-06 | 1E-06 | 0.0003 | 0.0046 | 2E-07 | 1E-06 | 1E-06 | 3E-05 | 7E-06 |
| Mathematics | 0.0003 | 0.0001 | 3E-05 | 0.0003 | 0.0002 | 0.0001 | 0.0001 | 0.0005 | 0.0002 | 0.0002 | 0.0002 | 1E-05 | 0.0006 | 3E-05 | 3E-05 | 0.0002 | 0.0309 | 0.0002 | 6E-05 | 1E-06 | 3E-05 | 0.0009 | 8E-05 | 3E-07 | 9E-08 | 7E-06 |
| Medicine | 0.0047 | 0.0021 | 9E-05 | 0.0122 | 6E-05 | 0.0001 | 0.0004 | 0.0002 | 1E-05 | 7E-05 | 1E-04 | 5E-05 | 0.0004 | 0.0005 | 0.0073 | 9E-05 | 0.0001 | 0.1705 | 0.0045 | 0.0025 | 0.0024 | 0.0001 | 0.0016 | 0.0008 | 0.0001 | 0.0002 | 0.0008 |
| Neuroscience | 0.0016 | 0.0006 | 6E-05 | 0.0044 | 3E-06 | 2E-05 | 5E-05 | 9E-05 | 4E-07 | 2E-06 | 7E-06 | 2E-07 | 8E-05 | 6E-05 | 0.0004 | 5E-06 | 2E-05 | 0.0038 | 0.0212 | 2E-05 | 0.0009 | 4E-05 | 0.0004 | 2E-05 | 5E-06 | 2E-06 | 5E-05 |
| Nursing | 1E-05 | 0.0001 | 3E-06 | 0.0007 | 7E-06 | 4E-05 | 8E-06 | 4E-06 | 1E-07 | 2E-08 | 1E-05 | 4E-07 | 5E-05 | 1E-05 | 6E-05 | 3E-07 | 2E-06 | 0.0027 | 2E-05 | 0.0024 | 7E-05 | 2E-07 | 4E-05 | 6E-05 | 9E-07 | 2E-06 | 4E-05 |
| Pharmacology, Toxicology and Pharmaceutics | 0.0005 | 0.0005 | 8E-07 | 0.0034 | 3E-06 | 0.0001 | 0.0005 | 3E-05 | 3E-08 | 3E-05 | 6E-06 | 2E-05 | 0.0002 | 0.0003 | 0.0005 | 2E-05 | 1E-05 | 0.0028 | 0.001 | 0.0083 | 6E-05 | 2E-05 | 2E-05 | 6E-06 | 2E-06 | 3E-05 |
| Physics and Astronomy | 0.0017 | 9E-05 | 1E-05 | 0.0006 | 3E-06 | 0.0011 | 0.0042 | 0.0006 | 1E-05 | 0.0008 | 4E-06 | 0.0003 | 0.0041 | 9E-05 | 9E-05 | 0.0041 | 0.0006 | 0.0001 | 5E-05 | 7E-07 | 7E-05 | 0.0785 | 5E-06 | 6E-06 | 8E-07 | 1E-06 | 2E-05 |
| Psychology | 9E-05 | 0.0001 | 0.0002 | 0.0003 | 0.0003 | 5E-06 | 2E-06 | 2E-05 | 4E-06 | 2E-06 | 4E-05 | 2E-06 | 3E-05 | 1E-05 | 1E-05 | 1E-07 | 1E-05 | 0.0017 | 0.0006 | 3E-05 | 7E-05 | 5E-06 | 0.0074 | 0.0006 | 1E-06 | 5E-07 | 2E-05 |
| Social Sciences | 8E-05 | 0.0002 | 0.0002 | 0.0001 | 0.0005 | 2E-05 | 3E-05 | 0.0002 | 2E-05 | 7E-05 | 0.0005 | 7E-05 | 0.0002 | 0.0002 | 7E-05 | 1E-05 | 7E-05 | 0.001 | 6E-05 | 5E-05 | 3E-05 | 8E-06 | 0.0007 | 0.0148 | 3E-06 | 2E-06 | 1E-05 |
| Veterinary | 2E-05 | 0.0001 | 7E-08 | 0.0003 | 2E-05 | 2E-06 | 6E-06 | 1E-06 | 6E-08 | 1E-06 | 2E-08 | 1E-07 | 3E-06 | 3E-05 | 0.0002 | 1E-06 | 3E-07 | 0.0003 | 2E-05 | 3E-06 | 4E-05 | 1E-06 | 3E-06 | 3E-05 | 0.0026 | 9E-07 | 4E-06 |
| Dentistry | 6E-06 | 5E-06 | 6E-08 | 9E-05 | 4E-09 | 2E-06 | 2E-06 | 4E-07 | 3E-08 | 1E-08 | 1E-08 | 1E-08 | 4E-06 | 1E-06 | 3E-05 | 1E-07 | 0.0003 | 5E-06 | 1E-06 | 9E-06 | 1E-06 | 7E-07 | 1E-06 | 2E-07 | 0.0026 | 6E-07 |
| Health Professions | 1E-05 | 3E-05 | 2E-05 | 0.0002 | 2E-06 | 2E-06 | 2E-05 | 1E-05 | 8E-07 | 2E-07 | 1E-05 | 2E-05 | 3E-05 | 4E-06 | 2E-05 | 9E-07 | 3E-06 | 0.0008 | 8E-05 | 6E-05 | 3E-05 | 2E-05 | 3E-05 | 1E-05 | 4E-07 | 6E-07 | 0.0023 |

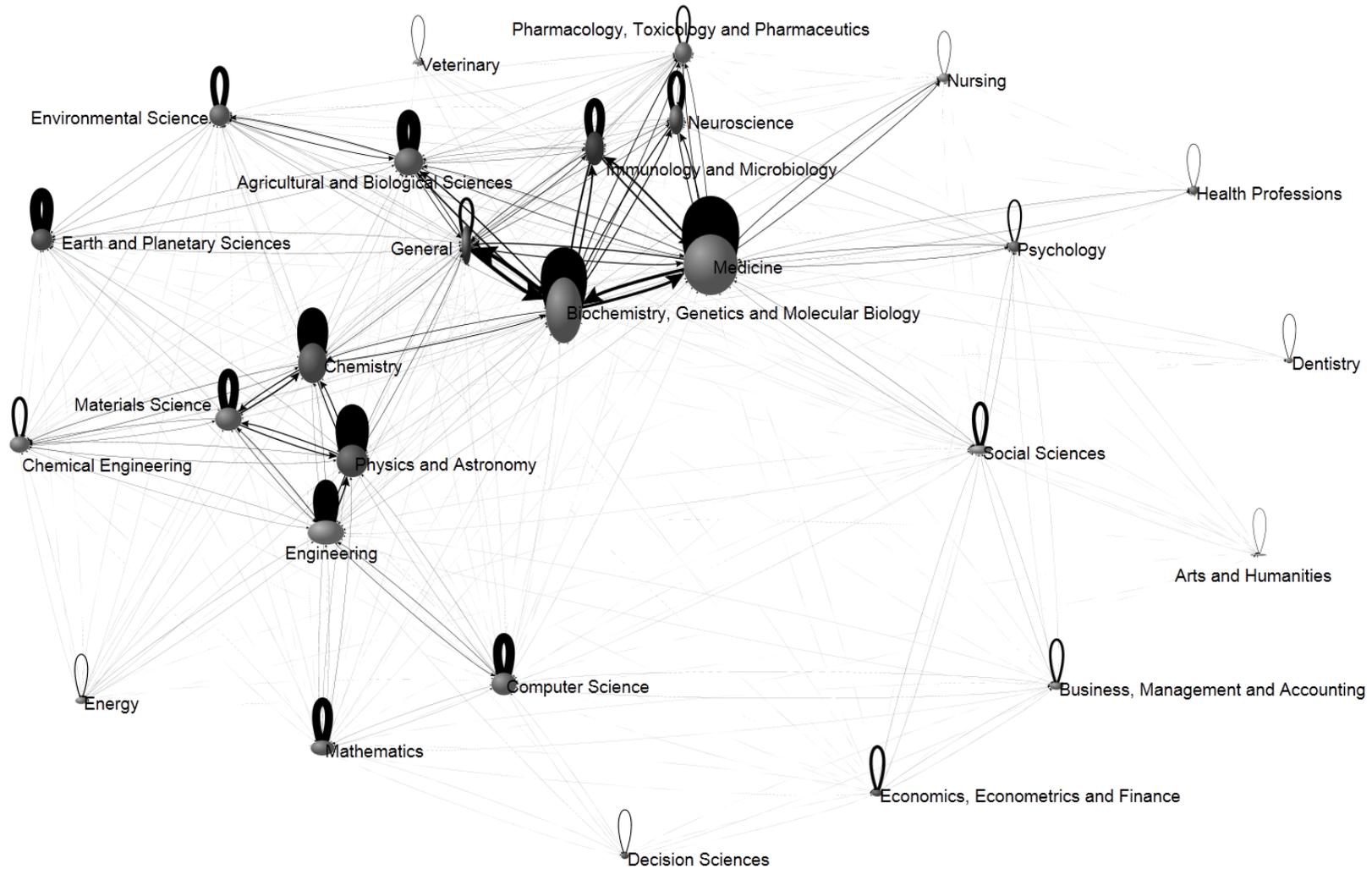

Figure 4: Network formed by the Prestige SJR2 transferred between Subject Areas. The width of each node is proportional to the number of documents, the height to the citations received, and the grey scale to the accumulated Prestige SJR2. The thickness of the links is proportional to the Prestige SJR2 transferred.

## 5. Conclusions

Beyond the metrics of the prestige of scientific journals which weight the Citation in terms of the prestige of the citing journal, the present SJR2 indicator solves the problem of the tendency for prestige scores to decrease over time by the use of stochastic matrices. It endows the resulting scores with meaning, and uses the cosine between the cocitation profiles of the citing and cited journals to weight the thematic relationship between the two journals.

The problem of the tendency for the scores to decrease as the calculation incorporates ever more journals and documents is overcome by dividing a journal's portion of prestige gained by the portion of citable documents. This means that if the journal is precisely at the mean, the two portions will be the same and the score will be unity. A higher score will mean that the portion of prestige is greater than that of citable documents, and vice versa. At the same time, this makes the weighted average of the scores obtained by the journals remain constant and equal to unity for every year, regardless of the number of scientific journals or documents counted in the calculation.

Using the cosine of the cocitation profiles is equivalent to assigning greater weight to citations to thematically close journals. For example, it increases the weight of citations to journals in the same Subject Area, and especially in the same Specific Subject Area. On the contrary, it decreases the weight of citations to scientific journals in other areas in which one must presume that the citing journal is of less authority. This leads to greatly equalizing the distribution by Subject Area, and especially by Specific Subject Area, and makes scores from different areas more comparable, all without using any arbitrary classification of journals or weights to apply to the citations.

While the resulting indicator has high Pearson and Spearman coefficients of correlation with the SNIP and JIF metrics overall, and by Subject Area and Specific

Subject Area, in our opinion it represents a step forward towards the best representation of the real prestige of scientific journals.

## 6. Acknowledgments

This work was financed by the Junta de Extremadura e Consejería de Educación Ciencia & Tecnología and the Fondo Social Europeo as part of Research Group grant GR10019, and by the Plan Nacional de Investigación Científica, Desarrollo e Innovación Tecnológica 2008e2011 and the Fondo Europeo de Desarrollo Regional (FEDER) as part of research projects TIN2008-06514-C02-01 and TIN2008-06514-C02-02.

## 7. References

Bar-Ilan, J. (2008) Which h-index? — A comparison of WoS, Scopus and Google Scholar. Scientometrics 74(2), 257-271

Bergstrom, C. (2007) Eigenfactor: Measuring the value and prestige of scholarly journals. College & Research Libraries News 68(5), 314–316.

Bollen, J., Rodriguez, M.A. & van de Sompel, H. (2006). Journal status, Scientometrics, 69(3), 669–687.

Bonacich, P. (1987), Power and centrality: A family of measures. American Journal of Sociology, 92(5), 1170–1182.

Brooks, T.A. (1985). Private acts and public objects: an investigation of citer motivations. Journal of the American Society for Information Science, 36(4): 223-229.

Garfield, E. (2006). The history and meaning of the journal impact factor. JAMA-Journal of the American Medical Association, 295(1), 90-93.

González-Pereira, B., Guerrero-Bote, V. P., & Moya-Anegón, F. (2010). A new approach to the metric of journals ' scientific prestige : The SJR indicator. Journal of Informetrics, 4(3), 379-391. doi: 10.1016/j.joi.2010.03.002.

Guerrero-Bote, V. P., Zapico-Alonso, F., Espinosa-Calvo, M. E., Gómez-Crisóstomo, R., & Moya-Anegón, F. (2007). The Iceberg Hypothesis: Import-Export of Knowledge between scientific subject categories. Scientometrics 71(3): 423-441.

Jacso, P. (2009). Péter's Digital Reference Shelf. Scopus. Available at: http://www.gale.cengage.com/reference/peter/200906/scopus.html. [Accessed: 1 October 2009]

Kostoff, R. N. (1997). The principles and practices of peer review. Science and Engineering Ethics, 3: 19-34.

Moed, H.F. (2005). Citation analysis in research evaluation. Dordrecht; Springer, p. 346.

Moed, H.F. (2010). Measuring contextual citation impact of scientific journals. Journal of Informetrics, 4(3), 265–277.

Laguardia, C. (2005), E-views and reviews: Scopus vs. Web of Science. Library Journal, 15. Available at: http://www.libraryjournal.com/article/CA491154.html%2522. [Accessed: 1 October 2009]

Lancho-Barrantes, B.S., Guerrero-Bote, V.P., & Moya-Anegón, F. (2010a). The iceberg hypothesis revisited. Scientometrics, 85 (2) 443-461.

Lancho-Barrantes, B.S., Guerrero-Bote, V.P., & Moya-Anegón, F. (2010b). What lies behind the averages and significance of citation indicators in different disciplines? Journal of Information Science. 36 (3), 371-382.

Leydesdorff, L., Moya-Anegón, F. & Guerrero-Bote, V.P. (2010). Journal maps on the basis of Scopus data: A comparison with the Journal Citation Reports of the ISI. Journal of the American Society for Information Science and Technology, 61 (2), 352-369.

Janssens, F., Zhang, L., Moor, B. & Glänzel, W. (2009). Hybrid clustering for validation and improvement of subject-classification schemes. Information Processing and Management 45, 683–702.

Lundberg, J. (2007). Lifting the crown—citation z-score. Journal of Informetrics, 1, 145–154 .

Marshakova, I. V. (1973) System of document connection based on references, Nauchno-Teknichescaya Informatisya, Series II (6): 3-8.

McCain, K. W. (1991). Mapping economics through the journal literature: An experiment in journal cocitation analysis. Journal of the American Society for Information Science, 42(4), 290-296. doi: 10.1002/(SICI)1097-4571(199105)42:4<290::AID-ASI5>3.0.CO;2-9.

Moed, H.F. (2005). Citation Analysis in research evaluation. Dordrecht; Springer, p. 346.

Moya-Anegón, F., Chinchilla-Rodríguez, Z., Vargas-Quesada, B., Corera-Álvarez, E., Muñoz-Fernández, F. J., González-Molina, A. & Herrero-Solana, V. (2007). Coverage analysis of Scopus: A journal metric approach. Scientometrics, 73, (1) , 53-78.

Moya-Anegón, F., Vargas-Quesada, B., Herrero-Solana, V., Chinchilla-Rodríguez, Z., Corera-Álvarez, E., & Muñoz-Fernández, F. J. (2004). A new technique for building maps of large scientific domains based on the cocitation of classes and categories. Scientometrics, 61(1), 129-145. doi: 10.1023/B:SCIE.0000037368.31217.34.

Page, L., Brin, S., Motwani, R. & Winograd, T. (1998). The PageRank citation ranking: Bringing order to the Web. Technical report, Stanford University, Stanford, CA, 1998.

Palacios-Huerta, I & Volij, O. (2004). The measurement of intellectual influence. Econometrica, 72(3), 963-977.

Pinski, G. & Narin, F. (1976), Citation influence for journal aggregates of scientific publications: Theory, with application to the literature of physics. Information Processing and Management, 12, 297-312.

SCImago Journal and Country Rank. SCImago Research Group. Available at: http://www.scimagojr.com [Accessed: 20 October 2011].

Small, H. (1973). Co-citation in the scientific literature: a new measure of the relationship between two documents, Journal of the American Society for Information Science (JASIS), 24, 265-269.

White, H. D., Mccain, K. W. (1998). Visualizing a discipline: an author co-citation analysis of information science, 1972-1995, Journal of the American Society for Information Science (JASIS), 49, 327-355.